\documentclass[twocolumn,aps,superscriptaddress]{revtex4-1}

\usepackage{extarrows}
\usepackage{amsmath}
\usepackage{graphicx}
\usepackage{bm}

\newcommand{\bracket}[1]{\langle#1\rangle}
\newcommand{\ket}[1]{|#1\rangle}
\newcommand{\eps}{\varepsilon}
\DeclareMathOperator{\Tr}{Tr}
\DeclareMathOperator{\RE}{Re}
\DeclareMathOperator{\IM}{Im}

\usepackage[colorlinks=true, letterpaper=true, pdfstartview=FitV,
  linkcolor=blue, citecolor=blue, urlcolor=blue]{hyperref}

\begin{document}

\title{Microscopic Theory of Spin Toroidization in Periodic Crystals}

\author{Yang Gao}

\affiliation{Department of Physics, Carnegie Mellon University,
  Pittsburgh, PA 15213, USA}

\author{David Vanderbilt}

\affiliation{Department of Physics and Astronomy, Rutgers University,
  Piscataway, New Jersey 08854-8019, USA}

\author{Di Xiao}

\affiliation{Department of Physics, Carnegie Mellon University,
  Pittsburgh, PA 15213, USA}

\date{\today}

\begin{abstract}
Using the semiclassical theory of electron dynamics, we derive a
gauge-invariant expression for the spin toroidization in a periodical
crystal.  We show that the spin toroidization is comprised of two
contributions: one is due to the configuration of a classical spin
array, while the other comes from the coordinate shift of the electron
as spin carrier in response to the inhomogeneous magnetic field. We then establish a direct and elengant relation
between our spin toroidization and the antisymmetric magnetoelectric
polarizability
in insulators. Finally, we demonstrate our spin toroidization in a tight-binding model and show that it is a genuine bulk quantity.
\end{abstract}

\maketitle

There has been continuous interest in
toroidal moments in
crystals~\cite{Dubovik1990,Gorbatsevich1994,Ederer2007,Spaldin2008},
mainly due to their intriguing role in various magnetoelectric
effects~\cite{Fiebig2005,Arima2005,Sawada2005,Aken2007,Zimmermann2014,Hayami2014,Fiebig2016}.
A toroidal moment is generally associated with a vortex-like
structure of magnetic moments.  Its spontaneous ordering characterizes
a ferrotoroidal state that
may exhibit a non-vanishing magnetoelectric
effect.  The density of the toroidal moment, called toroidization,
also constitutes an essential building block in the free energy
expansion in inhomogeneous fields.  However, despite its importance, a
microscopic theory of the toroidization based on quantum mechanical
wave functions is still missing.

In crystals the toroidization can arise from two sources, the orbital
and spin moments.  Here we will focus only on the contribution from
spins~\cite{foot1}.  By treating the spins as classical vectors, it has been
proposed that the spin toroidization can be written
as~\cite{Ederer2007,Spaldin2008}
\begin{equation}\label{eq_toroidp}
\bm {\mathcal T} = \frac{g\mu_B}{2\hbar V} \sum_i \bm r_i \times \bm s_i \;,
\end{equation}
where $g$ is the gyromagnetic factor, $\mu_B$ is the Bohr magneton,
$\bm r_i$ and $\bm s_i$ are the position and spin of each lattice
site, and $V$ is the volume of the sample.  There is also a recent
attempt to obtain a microscopic theory of the spin toroidization by
treating $\bm r$ and $\bm s$ as operators and directly evaluating the
expectation of Eq.~\eqref{eq_toroidp} using Wannier
functions~\cite{Thole2016}.  However, the resulting expression is not
gauge-invariant.

In this work, we develop a quantum theory of spin toroidization in
crystals.  Using the semiclassical theory of electron
dynamics~\cite{Sundaram1999,Xiao2010}, we obtain a gauge-invariant
expression for the spin toroidization in terms of bulk Bloch
functions, which is amenable to implementation in first-principles
codes.  By considering the molecular insulator limit, we find that the
contributions to the spin toroidization consists of two parts with
clear physical interpretations: one is due to the configuration of a
classical spin array, similar to Eq.~\eqref{eq_toroidp}, while the
other comes from the coordinate shift of the electron as spin carrier
in response to the inhomogeneous magnetic field.

Using our theory, we are able to establish a direct and elegant
relation between the spin toroidization and the antisymmetric
magnetoelectric polarizability in the case of insulators [see
  Eq.~\eqref{eq_connt}].  Such a relation is dictated by general
thermodynamic principles.  Finally, using a tight-binding toy model, we show that our spin toroidization is a genuine bulk quantity.

\section{General Formalism}

In this section, we first introduce the definition of the toroidization as a response function of the free energy to the derivative of the magnetic field. Then we use the semiclassical theory to derive the spin toroidization.
\subsection{Toroidization as a response function}
Our starting point
is the free energy density $F(\bm r)$ in an inhomogeneous magnetic
field $\bm B(\bm r)$.  Suppose that $\bm B(\bm r)$ is small and varies
slowly in space.  At a given point $\bm r$, we can perform a gradient
expansion of $F(\bm r)$ up to first order with respect to the
derivatives of the magnetic field
\begin{equation} \label{eq_exp}
\begin{split}
F(\bm r) &= F_0(\bm r) - \bm M \cdot \bm B(\bm r) -\mathcal{Q}_{ij} \partial_iB_j(\bm r)
+ \cdots \;,
\end{split}
\end{equation}
where $F_0(\bm r)$ is the free energy density at $\bm B=0$ and $\bm M$
is the magnetization. Here and hereafter the Einstein summation
convention is implied for repeated indices. The quantity $\mathcal{Q}_{ij}$ is the magnetic quadrupole moment density. The toroidization $\bm {\mathcal T}$ is the antisymmetric part of $\mathcal{Q}_{ij}$:
\begin{equation}\label{eq_tq}
\mathcal{T}_k=\frac{1}{2} \epsilon_{ijk}\mathcal{Q}_{ij}\,,
\end{equation}
where $\epsilon_{ijk}$ is the total antisymmetric tensor.

Based on Eq.~\eqref{eq_exp}, we can obtain a linear-response
expression for $\bm {\mathcal T}(\bm{r})$ by treating $\bm B(\bm r)$
and $\bm \nabla\times \bm B(\bm r)$ as independent variables,
arriving at
\begin{equation}\label{eq_diff}
\bm {\mathcal T}(\bm r) = -\lim_{\bm B(\bm r)\to 0}
\frac{\partial F(\bm r)}{ \partial(\bm \nabla\times \bm B)}
\biggr|_{\bm B(\bm r)} \;.
\end{equation}
Here the subscript $\bm B(\bm r)$ in taking the derivative with
respect to $\bm \nabla\times \bm B$ means that the magnetic field at the
point $\bm r$ has to be kept fixed as $\bm \nabla\times \bm B(\bm r)$ is
varied.

\subsection{Semiclassical theory of the spin toroidization}
 With the above definition of the toroidization, we now formulate its
microscopic theory. We will focus only on the spin toroidization.
Therefore, we take $\bm B(\bm r)$ as the Zeeman field, which couples
to the spin operator $\hat{\bm s}$. Then the full Hamiltonian can be
written as
\begin{equation} \label{eq_hamil}
\hat{H}_\text{F}=\hat{H}(i\hbar \partial_{\bm r},\bm r)
-\frac{g\mu_B}{\hbar} \bm B(\bm r)\cdot \hat{\bm s}\;.
\end{equation}
For definiteness we can consider the context to be that of a
spinor implementation of density functional theory with spin-orbit
interactions included.  The first term, $\hat{H}(i\hbar\partial_{\bm r},\bm r)$,
describes a perfect
crystal in the absence of a Zeeman field, while
the second term is inhomogeneous and breaks the translational
symmetry, making it difficult to diagonalize the
Hamiltonian~\eqref{eq_hamil} analytically.  Here we take a different
route by using the semiclassical theory of electron
dynamics~\cite{Sundaram1999,Xiao2010}, which is designed to study
Bloch electrons subject to perturbations varying slowly in
space.

In the spirit of the semiclassical theory, each Bloch electron
responds to the external Zeeman field in the form of a wave packet,
which has a specified
center of mass position $\bm r_c$ and momentum
$\bm k_c$. To construct the wave packet, we make a local approximation
and assume that the system can be described by a set of local
Hamiltonians $\hat{H}_c[\bm B(\bm r_c)]=\hat{H}(i\hbar\partial_{\bm
  r},\bm r)-(g\mu_B/\hbar)\bm B(\bm r_c)\cdot \bm s$. Since
$\hat{H}_c$ respects the lattice translational symmetry, its
eigenstate has the form of a Bloch function $e^{i\bm k\cdot \bm r}
\ket{\tilde{u}_n(\bm k,\bm B(\bm r_c))}$ with the eigenenergy
$\tilde{\eps}(\bm k_c,\bm B(\bm r_c))$, where $n$ is the band
index. In the limit $\bm B(\bm r_c)\to 0$, $\ket{\tilde{u}_n(\bm
k,\bm B(\bm r_c))}$ reduces to $\ket{u_n(\bm k)}$, the
periodic part of the Bloch function of $\hat{H}$, and
$\tilde{\eps}_n(\bm k_c,\bm B(\bm r_c))$ reduces to
$\eps_n(\bm k_c)$, the eigenenergy of the unperturbed
Hamiltonian $\hat{H}$.  For simplicity, we hereafter drop the argument
of $\ket{\tilde{u}_n}$, $\ket{u_n}$, $\tilde{\eps}_n$ and
$\eps_n$.
For illustrative purposes, we consider a single band
with index $0$, and the wave packet is thus the superposition of
$e^{i\bm k\cdot \bm r} \ket{\tilde{u}_0}$.

The wave-packet dynamics can be properly formulated as a set of
semiclassical equations of motion in the phase space spanned by $\bm
r_c$ and $\bm k_c$~\cite{Sundaram1999,Xiao2010}. The spatial
inhomogeneity of $\bm B(\bm r)$ introduces two
essential ingredients for the purpose of evaluating the spin
toroidization in Eq.~\eqref{eq_diff}.  First, the phase space density
of states $\mathcal{D}$ is modified.  It has the form~\cite{Xiao2005} (see also Sec.~VI.B of Ref.~\onlinecite{Xiao2010})
\begin{equation}
\mathcal{D}(\bm r_c,\bm k_c) = 1 + \Tr(\Omega_{\bm k,\bm r}) \;,
\end{equation}
where
\begin{equation}
(\Omega_{\bm k,\bm r})_{ij}=-2\IM\bracket{\partial_{k_{ci}}\tilde{u}_0|\partial_{r_{cj}}\tilde{u}_0}
\end{equation}
is the mixed Berry curvature between the real and momentum space.
This modified density of states has been applied to derive
the polarization in inhomogeneous crystals~\cite{Xiao2009}.
Secondly, the band energy $\tilde{\eps}$ is also affected by
the spatial inhomogeneity~\cite{Sundaram1999}
\begin{equation}
\eps_0^\prime=\tilde{\eps}_0 + \IM \bracket{\partial_{k_{ci}}\tilde{u}_0| (\tilde{\eps}_0-\hat{H}_c)|\partial_{r_{ci}} \tilde{u}_0} \;.
\end{equation}

With the above two ingredients we are ready to evaluate the free
energy density $F$.  For simplicity we set $T = 0$.  The free energy
density is given by $F=\int \frac{d\bm k_c}{ (2\pi)^3} \mathcal{D}(\bm
r_c,\bm k_c)
(\eps_0^\prime-\mu)\Theta(\mu-\eps_0^\prime)$, where
$\Theta$ is the Heaviside function. At first order with respect to the
derivative of $\bm B$, the correction to the free energy density is
\begin{equation}\label{eq_freeeng}
\delta F=-\int^\mu \frac{d\bm k_c}{(2\pi)^3} \IM\bracket{\partial_{k_{ci}}\tilde{u}_0| (\tilde{\eps}_0+\hat{H}_c-2\mu)|\partial_{r_{ci}} \tilde{u}_0}\;.
\end{equation}
Here the upper limit $\mu$ means that the integration is taken up to
$\eps_0=\mu$.

The toroidization defined in Eq.~\eqref{eq_diff} can be
obtained from the above free-energy correction. Since
$\ket{\tilde{u}_0}$ depends on $\bm r_c$ through $\bm B$, we make
the substitution
$\partial_{r_{ci}}\ket{\tilde{u}_0}=\partial_{r_{ci}}B_\ell
\partial_{B_\ell}\ket{\tilde{u}_0}$. We then collect terms involving
the antisymmetric part of $\partial_{r_{ci}}B_\ell$ and take the
derivative as in Eq.~\eqref{eq_diff}. The final expression is
\begin{align}\label{eq_tor1}
\bm {\mathcal T}=\frac{1}{ 2} \int^\mu \frac{d\bm k}{ (2\pi)^3} \IM \bracket{\partial_{\bm k}\tilde{u}_0| \times(\tilde{\eps}_0+\hat{H}_c-2\mu)|\partial_{\bm B} \tilde{u}_0} \Bigr|_{\bm B\to 0}\;.
\end{align}
Here and hereafter we drop the subscript $c$ of $\bm k_c$.
Note that Eq.~\eqref{eq_tor1} can be straightforwardly
generalized to the multiband case by summing over all occupied states [see Eq.~\eqref{TT}],
but we continue to focus on the single-band case here.

The structural similarity between Eq.~\eqref{eq_tor1} and the orbital
magnetization formula~\cite{Xiao2005,Thonhauser2005,Ceresoli2006,Shi2007} is
striking.  In fact,
by making the substitution $\bm \partial_{\bm B}\to \bm
\partial_{\bm k}$, Eq.~\eqref{eq_tor1} exactly coincides with the expression of
the orbital magnetization. This similarity has its root in the nature of
spin toroidization and orbital magnetization: they both measure
the moment of some observable, which is spin for spin
toroidization and velocity for orbital magnetization.

Equation~\eqref{eq_tor1} can be cast in a form involving only
unperturbed Bloch states $|u_0\rangle$ instead of
$|\tilde{u}_0\rangle$.  Using the perturbation theory,
up to the first order in the Zeeman field we have
\begin{equation}
\ket{\tilde{u}_0}=\ket{u_0}-\frac{g\mu_B}{\hbar} \sum_{n\neq 0}\frac{\bm B\cdot \bm s_{n0}}{ \eps_0-\eps_n} \ket{u_n}\;.
\end{equation}
Then Eq.~\eqref{eq_tor1} can be rewritten as
\begin{equation}\label{eq_dft}
\bm {\mathcal T}=-\frac{g\mu_B}{ 2}\sum_{n\neq 0}\int^\mu \frac{d\bm k}{ (2\pi)^3}(\eps_0+\eps_n-2\mu)\frac{\IM(\bm v_{0n}\times \bm s_{n0})}{ (\eps_0-\eps_n)^2}\;,
\end{equation}
where $\bm v_{0n}=\bracket{u_0|\hat{\bm v}|u_n}$ and
$\bm s_{n0}=\bracket{u_n|\hat{\bm s}|u_0}$ are the
interband elements of the velocity and spin operators, respectively.
Both Eq.~\eqref{eq_tor1} and \eqref{eq_dft} are amenable to implementation
in a first-principles calculation. To further check the validity of our result, we have also
carried out a linear response calculation~(see Appendix A for details), similar to the
derivation of the orbital magnetization in Ref.~\onlinecite{Shi2007}, and
obtained the same result.

We comment that in general the spin magnetic quadrupole moment density can obtained in a similar way. The result reads
\begin{equation}\label{eq_quad}
\mathcal Q_{ij}=-{g\mu_B}\sum_{n\neq 0}\int^\mu \frac{d\bm k}{ (2\pi)^3}(\eps_0+\eps_n-2\mu)\frac{\IM[(v_i)_{0n} (s_j)_{n0}]}{ (\eps_0-\eps_n)^2}\;.
\end{equation}
One can easily check that $\bm {\mathcal T}$ and $\mathcal Q_{ij}$ satisfy Eq.~\eqref{eq_tq}.

It is clear that our expression~\eqref{eq_dft} for the spin
toroidization is gauge-invariant since it does not change
if an arbitrary phase factor is applied to
$|u_n\rangle$.  As a consequence, the spin toroidization does not have
any quantum of uncertainty, and it always vanishes for a system with
either time-reversal or inversion symmetry.  This is in sharp contrast
to both the electric polarization~\cite{Kingsmith1993,Resta1994} and
the previous theory of the spin
toroidization~\cite{Ederer2007,Thole2016}.

It is also worth mentioning that our toroidization cannot be used to
predict a surface magnetization density, unlike the electric
polarization, which has a definitive relation to the surface charge
density~\cite{Vanderbilt1993}.  This difference can be traced to the
fact that charge is conserved but spin is not.

\section{Interpretation of spin toroidization}
In this section, we explore the physical meaning of the spin toroidization in the Wannier representation and discuss the difference between our result and the classical definition of spin toroidization in Eq.~\eqref{eq_toroidp}. Finally, we show that the spin toroidization can be directly related to the spin magnetoelectric polarizability.
\subsection{Molecular Insulator Limit}
To shed light on the physical
meaning of the spin toroidization in Eq.~\eqref{eq_tor1}, we rewrite
it for an insulator using the Wannier function
representation.  We label
the Wannier function defined from the local Hamiltonian $\hat{H}_c$ by
$|w_{0}(\bm R,\bm B)\rangle$, with $0$ being the band index and $\bm
R$ being the lattice site. In this representation Eq.~\eqref{eq_tor1}
becomes~(see Appendix B for details)
\begin{widetext}
\begin{equation}\label{eq_wan}
\begin{split}
\bm {\mathcal T}&=\frac{1}{ V_\text{cell}}\RE\bracket{w_0(\bm B)|\bm r (\hat{H}_c-\mu)\times \bm \partial_{\bm B}|w_0(\bm B)}\Bigr|_{\bm B\to 0}-\frac{g\mu_B}{ 2\hbar V_\text{cell}}\bracket{ w_0(\bm B)|\bm r\times \hat{\bm s}|w_0(\bm B)}\Bigr|_{\bm B\to 0} \\
&\quad-\frac{g\mu_B}{ 2\hbar V_\text{cell}}\sum_{\bm R}\bracket{w_0(\bm B)|\bm r|w_{0} (\bm R,\bm B)}\Bigr|_{\bm B\to 0}\times\bracket{w_{0}(\bm R,\bm B)|\hat{\bm s}|w_0(\bm B)}\Bigr|_{\bm B\to 0} \;,
\end{split}
\end{equation}
\end{widetext}
where $\ket{w_0(\bm B)} =\ket{w_{0}(\bm R,\bm B)}$ with $\bm R=0$,
and $V_\text{cell}$ is the unit cell volume.

The meaning of Eq.~\eqref{eq_wan} can be clarified further by taking
the molecular insulator limit.  Since the spin toroidal moment arises from
a vortex-like arrangement of spins, there must be multiple atoms in a unit cell,
which we call a molecule.  The molecular insulator limit is then taken by letting the distance
between neighboring molecules go to infinity while the relative structure of each
molecule is unchanged.  In this limit, $\ket{w_{0}(\bm R,\bm B)}$ is just the energy eigenstate of the molecule, translated to sit in cell $\bm R$.  We will further assume that the system respects the combined time reversal and inversion symmetry such that $\bracket{w_0|\hat{\bm s}|w_0}$ vanishes.

In the molecular insulator limit Eq.~\eqref{eq_wan} consists of two
parts~(see Appendix C for details). The first part is
\begin{equation} \label{eq_wan1}
\bm {\mathcal T}_1 =\frac{g\mu_B}{2\hbar V_\text{cell}}
\bracket{w_0(\bm B)|\bm r\times \hat{\bm s}|w_0(\bm B)}\Bigr|_{\bm B\to 0}\;.
\end{equation}
It is clear that this term is due to the configuration of an array of
classical spins, similar to the equation appearing as
Eq.~\eqref{eq_toroidp} in the classical picture.
The second part, coming from the modified density of states, is
\begin{equation}\label{eq_wan2}
\bm {\mathcal T}_2=-\frac{1}{2V_\text{cell}} (\eps_0-\mu)
(\partial_{\bm B}\times \bar{\bm r})\Bigr|_{\bm B\to 0}\;,
\end{equation}
where $\eps_0$ refers to the molecular electronic energy levels
and $\bar{\bm r}=\langle w_0(\bm B)|\bm r|w_0(\bm B)\rangle$
is the electron position
under the external Zeeman
field.  Here $\eps_0-\mu$ is the free energy for state $0$.

The $\bm {\mathcal T}_2$ term can be intuitively understood as follows.
In the spirit of the
molecular insulator limit, if each molecule is simply a cluster of
classical spins, under an inhomogeneous magnetic field the spins on
each site can rotate but cannot move.  However, the spins are carried
by electrons, and the inhomogeneous Zeeman field will exert a spin
force on the electron. Therefore, the electron will
shift to a new equilibrium position due to the balance between the
spin force and the restoring force that binds electrons to ions.
The corresponding energy change gives rise to $\bm {\mathcal T}_2$.  In a
semiclassical picture, $\partial_{\bm B}\times\bar{\bm r}$ counts the
change of the number of electronic states within a volume element located at $\bm r$.

Equation~\eqref{eq_wan2} also provides a strong hint connecting the
toroidization with the magnetoelectric polarizability.
Taking the derivative with respect to $\mu$ in Eq.~\eqref{eq_wan2}
yields $(1/2V_\text{cell})\bm \partial_{\bm B}\times
\bm \bar{\bm r}$.
Since $\bar{\bm r}$ is proportional to the electric
polarization, its derivative with respect to the Zeeman field $\bm B$
is exactly the magnetoelectric polarizability.  We show below that this is a general relation born out from our theory.

\subsection{Connection to magnetoelectric polarizability}

It is well known that the
toroidization and the antisymmetric part of the magnetoelectric
polarizability transform in the same way under symmetry
operations~\cite{Spaldin2008}.  However, an explicit relation
between these two quantities has not previously been identified.
Here we show that for an
insulator, the spin toroidization admits a direct and elegant
connection to the spin magnetoelectric polarizability.

According to the modern theory of
polarization~\cite{Kingsmith1993,Resta1994}, as we vary the $j$-th
component of the Zeeman field, the change of the polarization is given
by
\begin{equation}
\Delta P_i=e\int \frac{d\bm k dB_j}{ (2\pi)^3} \IM\bracket{\partial_{k_i} \tilde{u}_0|\partial_{B_j}\tilde{u}_0}\;.
\end{equation}
Therefore, the magnetoelectric polarizability has the form~\cite{Essin2010}
\begin{equation}\label{eq_me1}
\alpha_{ij}=\frac{\partial P_i}{\partial B_j} \biggr|_{\bm B\to 0}
=e \int \frac{d\bm k}{(2\pi)^3} \IM
\bracket{ \partial_{k_i} \tilde{u}_0|\partial_{B_j}\tilde{u}_0}\Bigr|_{\bm B\to 0}\;.
\end{equation}
On the other hand, note that for an insulator the Fermi-surface
contribution to $\bm {\mathcal T}$ vanishes.  If we take the derivative of
Eq.~\eqref{eq_tor1} with respect to $\mu$, we find the
desired connection
\begin{equation}\label{eq_connt}
e\frac{\partial {\mathcal T}_k}{ \partial \mu}=- \frac{1}{2} \epsilon_{ijk}\alpha_{ij} \;,
\end{equation}
where $\epsilon_{ijk}$ is the Levi-Civita symbol.

There is a heuristic derivation of the relation~\eqref{eq_connt}.
Equation~\eqref{eq_exp} suggests that the differential form of the free energy is $dF = -\bm {\mathcal T} \cdot d(\bm\nabla \times \bm B) - \rho d\mu$, where $\rho$ is the particle density.  We can then obtain via the Maxwell relation
\begin{equation} \label{maxwell}
\frac{\partial {\mathcal T}_i}{\partial\mu} = \frac{\partial \rho}{\partial(\bm\nabla\times \bm B)_i} \;.
\end{equation}
For a given point $\bm r$,  we write $\bm B = (1/2)\bm h \times \bm r$ such that $\bm\nabla \times \bm B = \bm h$ and $\bm B$ vanishes exactly at $\bm r$.  This  choice ensures that $\partial_iB_j$ only has the antisymmetric component.  On the other hand, the application of an inhomogeneous Zeeman field $\bm B(\bm r)$ will induce an inhomogeneous polarization $\bm P(\bm r)$, which in turn leads to a charge density change, i.e.,
\begin{equation} \label{polar}
e\rho = \bm\nabla \cdot \bm P = \partial_i(\alpha_{ij} B_j)
= \frac{1}{2} \epsilon_{ijk} \alpha_{ij} h_k \;.
\end{equation}
Combining Eq.~\eqref{maxwell} and \eqref{polar} then yields Eq.~\eqref{eq_connt}.

In fact, similar relation exists between the spin magnetic quadrupole moment density $\mathcal Q_{ij}$ and each component of the spin magnetoelectric polarizability. Using similar derivations, one can find that
\begin{equation}
e\frac{\partial \mathcal Q_{ij}}{ \partial \mu}=- \alpha_{ij}\,.
\end{equation}
This relation is also implied from Eq.~\eqref{eq_connt} based on the definition~\eqref{eq_tq}.

We note that the above argument is thermodynamic in nature and does not depend on microscopic details.  Therefore, it is valid for any physical systems, and also for orbital toroidization, where the Zeeman field $\bm B$ is replaced by a magnetic field.

\section{Spin toroidization as a bulk quantity}

For a response function to reflect the bulk properties of the sample, it is essential that the response function has a valid thermodynamics limit, i.e., it has a well-defined limit as the sample size grows to infinity. This is well illustrated in the modern theory of electric polarization and orbital magnetization~\cite{Bianco2013,Marrazzo2016,Ceresoli2006,Thonhauser2005,Kingsmith1993,Vanderbilt1993,Shi2007,Resta1994}.

\begin{figure}[t]
\setlength{\abovecaptionskip}{0pt}
\setlength{\belowcaptionskip}{1pt}
\scalebox{0.25}{\includegraphics*{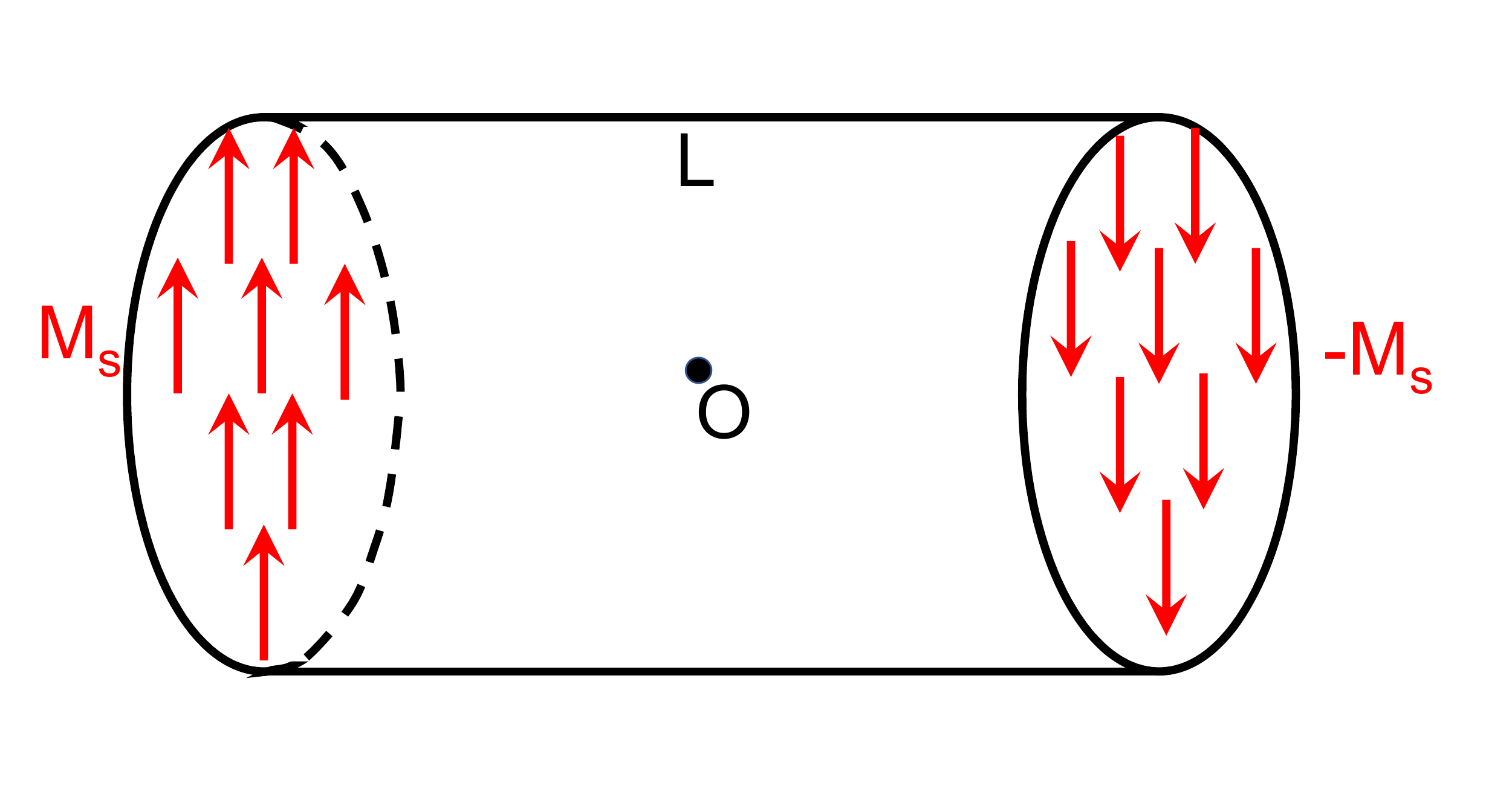}}
\caption{Sample with non-zero surface magnetization. Red arrows show the direction of the surface magnetization on the left and right surfaces.}
\label{fig_fig1}
\end{figure}

However, the spin toroidization in Eq.~\eqref{eq_toroidp} in the classical picture is not a bulk quantity. To show this, we consider the bulk sample in Fig.~\ref{fig_fig1}, which has uniform but opposite surface magnetization on the left and right surfaces. We choose the origin to be the center of the sample and label the left and right surface magnetization, the surface area, and the distance between two surfaces by $M_\text{s}$, $-M_\text{s}$, $S$, and $L$, respectively. Then from Eq.~(\ref{eq_toroidp}), the contribution from the surface reads $g\mu_B/2 [(L/2)SM_\text{s}/V+(L/2)SM_\text{s}/V]=g\mu_B M_\text{s}/2$, which obviously does not vanish as $V\rightarrow \infty$. Moreover, this contribution can point to any direction as the direction of the surface magnetization varies.

In comparison, our spin toroidization in Eq.~(\ref{eq_tor1}) is indeed a genuine bulk quantity. In the following, we demonstrate this point by studying a toy model.

\subsection{Toy model}

\begin{figure}[t]
\includegraphics[width=\columnwidth]{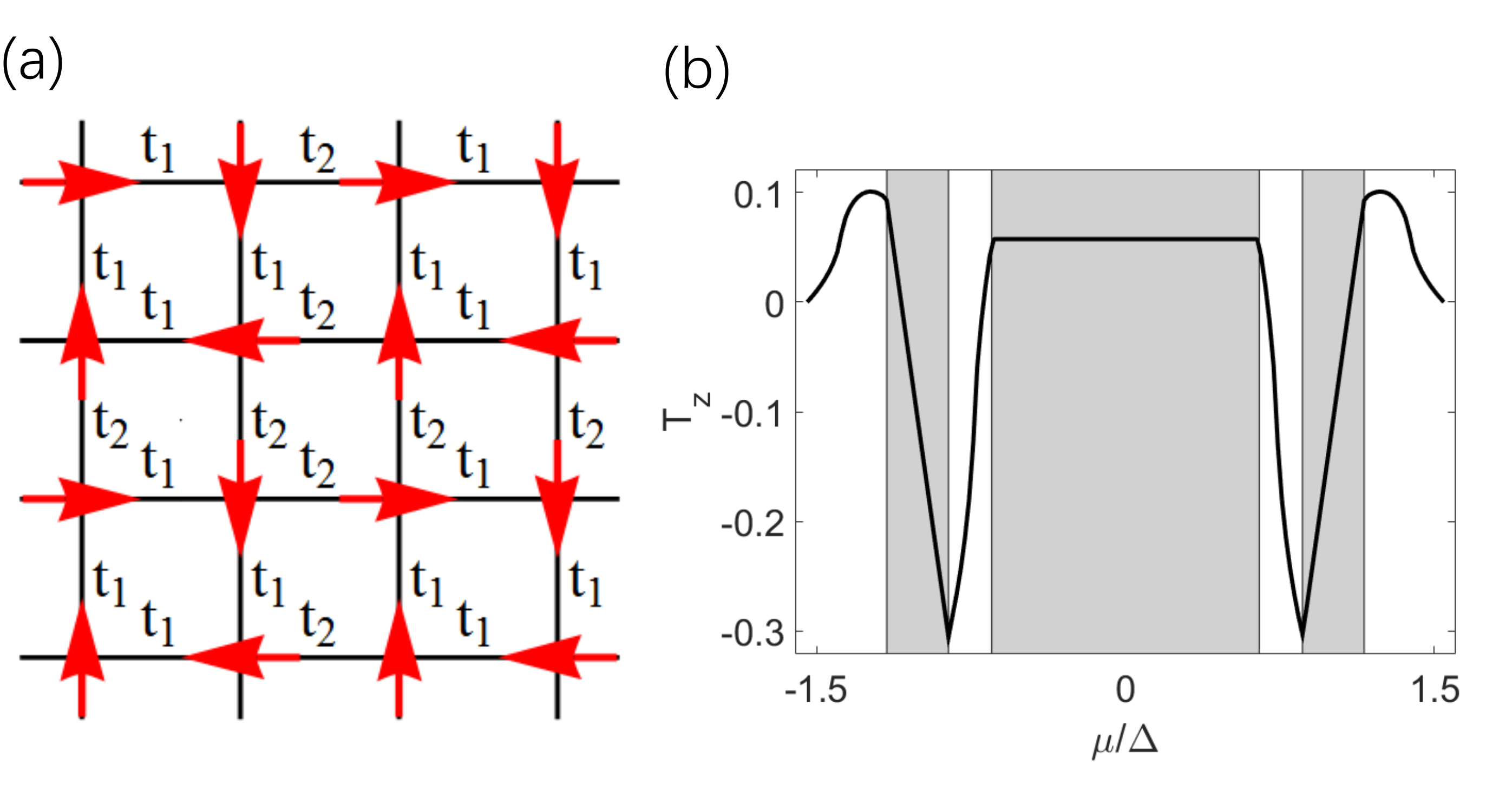}
\caption{Spin toroidization of a tight-binding model. Panel (a) is part
  of a periodic crystal. $t_1$ and
  $t_2$ is the nearest neighbour hopping strength. The red arrow on
  each lattice site indicates the direction of the local exchange
  field. The lattice constant is $a/2$. Panel (b) is the calculated
  spin toroidization (in units of $g\mu_B/4a$) as a function of the
  chemical potential.  The shaded areas correspond to energy gaps.
  The parameters are chosen as follows:
  $t_1=0.3\Delta$ and $t_2=0.15\Delta$.}
\label{fig_fig2}
\end{figure}

We consider the model Hamiltonian
\begin{equation}\label{eq_tb}
\hat{H}_\text{TB} = -\Delta \sum_i \bm n_i \cdot \bm \sigma_{\alpha\beta}c_{i\alpha}^\dagger c_{i\beta} + \sum_{\bracket{i,j}} t_{ij} c_{i\alpha}^\dagger c_{j\alpha}\;,
\end{equation}
where $\Delta$ is the local exchange field, $\bm n_i$ is the exchange field direction, $\alpha$ and $\beta$ label the spin
components, and $t_{ij}$ is the spin-independent nearest neighbor
hopping strength alternating between $t_1$ and $t_2$ as shown in Fig.~\ref{fig_fig2}a.

Since $\bm {\mathcal T}$ transforms as a vector, it is useful to first analyze
the symmetry of this system. The system has a 4-fold rotational
symmetry about the vertical axis. Therefore, the toroidization cannot
have any in-plane component. Moreover, if $t_1=t_2$, the system also
respects the combined symmetry of the mirror operation $\sigma_h$
(i.e., $z\to -z$) followed by a translation across the
diagonal direction. This requires that the out-of-plane component of
the toroidization vanishes. Both results have been verified in our
numerical calculations. We thus focus on the case $t_1\neq t_2$
for which an out-of-plane toroidization is expected.

Figure~\ref{fig_fig2}b shows the toroidization calculated from
Eq.~\eqref{eq_tor1} as a function of the chemical potential.
The system
has four bands separated by three global band gaps, and each band is doubly degenerate.  The curve is symmetric with respect to $\mu = 0$ because
of the particle-hole symmetry of our model. When the chemical potential falls inside the lowest and highest gap region, the toroidization varies linearly, and we confirm that its slope coincides with the magnetoelectric polarizability, consistent with our Eq.~\eqref{eq_connt}. When the chemical potential falls inside the middle gap region, the toroidzation is a nonzero constant while the magnetoelectric polarizability vanishes, in accordance with the particle-hole symmetry. Our toy model thus represents an interesting scenario with a vanishing magnetoelectric polarizability but a finite spin toroidization when the Fermi energy is in the middle gap.

\subsection{Irrelevance of the boundary}
Now we consider a finite sample with the model Hamiltonian in Eq.~\eqref{eq_tb} using open boundary conditions.  We label the $n$-th eigenstate by $E_n$ and the corresponding wave function by $|\psi_n\rangle$. Eq.~(\ref{eq_dft}) reduces to
\begin{equation}\label{eq_finite}
\bm {\mathcal{T}}=-\sum_{m,n}{E_m+E_n-2\mu\over E_m-E_n}(\bracket{\psi_m|\bm r|\psi_n}\times \bracket{\psi_n|\bm \sigma|\psi_m}+\text{c.c.})\,,
\end{equation}
where $m$ labels occupied states and $n$ labels unoccupied states, and $\bm {\mathcal T}$ is in units of $g\mu_B/4S$ with $S$ being the area of the sample.

\begin{figure}[t]
\setlength{\abovecaptionskip}{0pt}
\setlength{\belowcaptionskip}{1pt}
\scalebox{0.26}{\includegraphics*{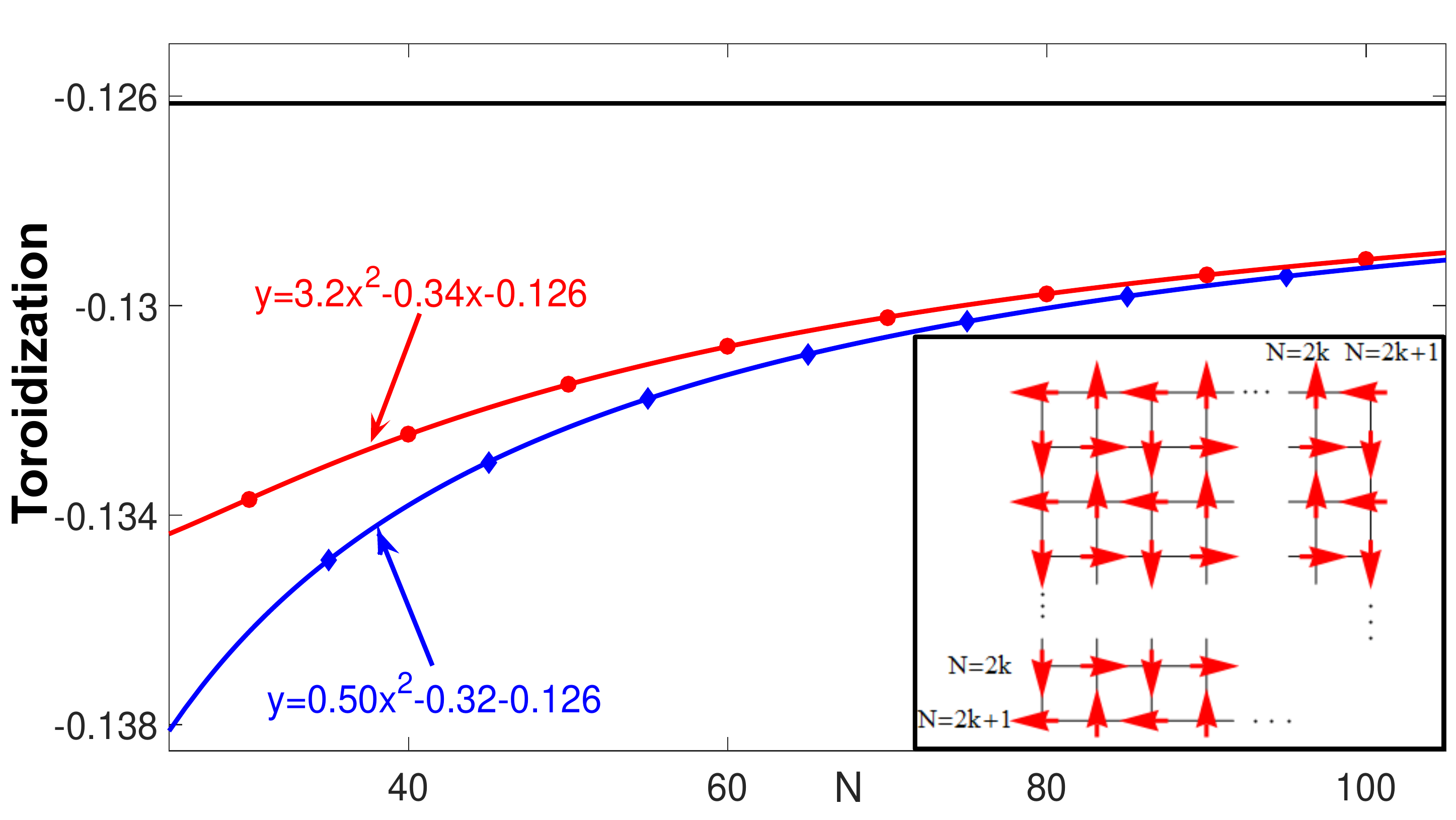}}
\caption{Toroidization in the bulk and in finite samples for $\mu=-0.9 \Delta$ (inside the band gap). The horizontal axis is the number of lattice sites $N$ on one edge; the vertical axis is the toroidization in units of $g\mu_B/4a$. The black line is the bulk value. The red and blue symbols are the finite-sample results for the toroidization, and their quadratic fittings based on Eq.~(\ref{eq_asy}) are displayed as the red and blue curves respectively. In the two equations, $x$ has the meaning of $1/N$. The inset illustrates the local exchange order in the finite sample, from which the hoppings can be derived via Fig.~\ref{fig_fig2}(a).}
\label{fig_fig3}
\end{figure}

The finite-sample value of the spin toroidization is
calculated based on Eq.~(\ref{eq_finite}) and plotted in
Fig.~\ref{fig_fig3}. The bulk value is $-0.126$ in units of
$g\mu_B/4a$ and is displayed as the straight black line in
Fig.~\ref{fig_fig3}. The red dots and blue diamonds are the finite-sample
results with an even and odd number of lattice sites along the sample edge
respectively. They fall on different curves due to the different
surface terminations
as shown in the inset. For
large $N$, the finite-sample results should asymptotically satisfy
\cite{Ceresoli2006}
\begin{equation}\label{eq_asy}
\mathcal T=\mathcal T_\text{bulk}+{a_1\over N}+{a_2\over N^2}
\end{equation}
where $\mathcal T_\text{bulk}$ is the bulk value. The second and third terms are due to the edge and corner contributions respectively. We find that the red and blue dots indeed fit Eq.~(\ref{eq_asy}) very well. This clearly demonstrates that our spin toroidization in Eq.~(\ref{eq_tor1}) and (\ref{eq_dft}) is a genuine bulk property.

\begin{figure}[t]
\setlength{\abovecaptionskip}{0pt}
\setlength{\belowcaptionskip}{1pt}
\scalebox{0.26}{\includegraphics*{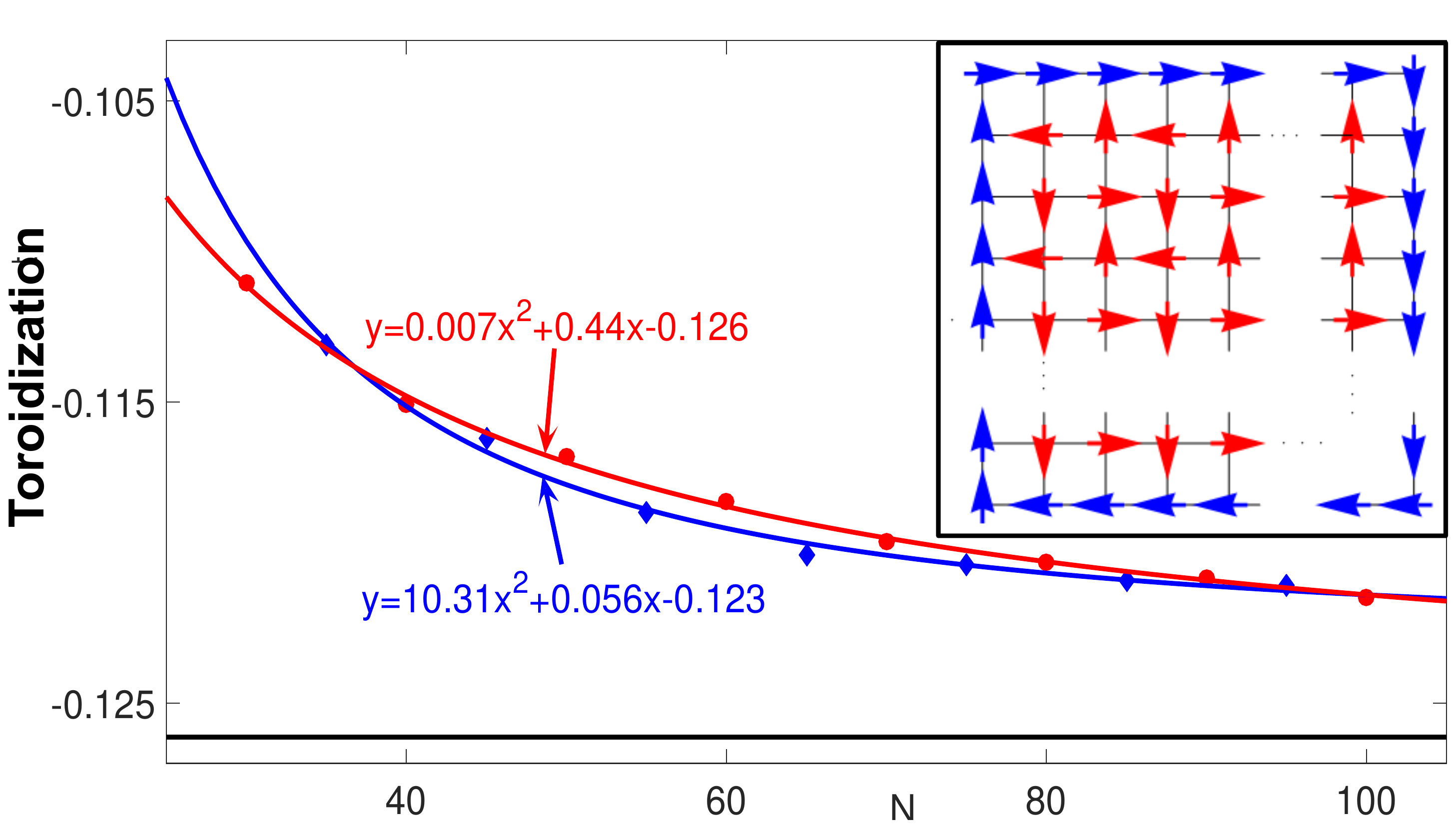}}
\caption{Toroidization for finite samples with surface perturbation. The axes and symbols have the same meaning as in Fig.~\ref{fig_fig3}. The inset shows the configuration of the finite sample. The layer with blue arrows is the additional perturbing layer. Here the original sample with red arrows has even $N$ along each edge. The odd $N$ case can be constructed in the same way. The hopping strength between the perturbation layer and the original sample is derived based on the known hopping strengths in the original sample in such a way that $t_{ij}$ still alternates between $t_1$ and $t_2$ for the whole sample.}
\label{fig_fig4}
\end{figure}

To further demonstrate the irrelevance of the boundary condition, we rotate the spin direction on the surface of our finite sample as shown in the inset of Fig.~\ref{fig_fig4}. This type of perturbation obviously contributes to Eq.~(\ref{eq_toroidp}) in the thermodynamic limit. In Fig.~\ref{fig_fig4}, we plot the calculated spin toroidization based on Eq.~(\ref{eq_dft}) for this type of finite samples. We find that although the toroidizations of finite samples are changed by the surface perturbation, they still follow the asymptotic rule in Eq.~(\ref{eq_asy}) very well and converge to the same bulk value.

\section{Summary}

In this work, we develop a quantum theory of spin toroidization in
crystals.  Using the semiclassical theory of electron
dynamics, we obtain a gauge-invariant
expression for the spin toroidization in terms of bulk Bloch
functions that is amenable to implementation in first-principles
codes. Using our theory, we are able to establish a direct and elegant
relation between the spin toroidization and the antisymmetric
magnetoelectric polarizability in the case of insulators, as dictated by general
thermodynamic principles.  Finally, using a tight-binding toy model, we show that our spin toroidization is a genuine bulk quantity.

\begin{acknowledgments}
This work is supported by the Department of Energy, Basic Energy Sciences,
Materials Sciences and Engineering Division, Grant No.~DE-SC0012509 (Y.G.\ and D.X.).
D.V.\ is supported
by the National Science Foundation, Grant No.~DMR-1408838.
\end{acknowledgments}

\begin{widetext}
\appendix
\section{Spin toroidization in linear response theory}

In this section we derive the spin toroidization using linear response theory, similar to the calculation of the orbital magnetization in Ref.~\onlinecite{Shi2007}. Without loss of generality, we only calculate the $z$-th component of the spin toroidization.

As given by Eq.~\eqref{eq_diff}, the spin toroidization is the response of the free energy density to an external Zeeman field.  Let us consider a Zeeman field of the following form
\begin{equation} \label{zeeman}
\bm B=(h/2q) (\sin(qx) \hat{y}-\sin(qy) \hat{x}) \;,
\end{equation}
where $h$ is small. This Zeeman field has a curl
\begin{equation}
(h/2)(\cos(qx)+\cos(qy)) \hat{z} \;,
\end{equation}
which reduces to $h$ in the limit $q\rightarrow 0$.  Note that the symmetric part of the derivative of the Zeeman field~\eqref{zeeman} is $\partial_x B_y+\partial_y B_x=(h/2)(\cos(qx)-\cos(qy))$, which vanishes in the limit $q\rightarrow 0$. As a result, the response of the free energy density to this Zeeman field in the limit $q\rightarrow 0$ is purely due to its curl.

At zero temperature, the free energy density reads $\hat{F}=\hat{H}-\mu \hat{N}$.  With the above Zeeman field, the change in $F$ can be divided into four parts:
\begin{equation}\label{suppl_eq_free}
\delta F(\bm r)=\sum_{n\bm k} (\delta f_{n\bm k}) \psi^*_{n\bm k} \hat{F}_0\psi_{n\bm k}+f_{n\bm k} \psi^*_{n\bm k} \bm B\cdot \hat{\bm s}\psi_{n\bm k}+f_{n\bm k}(\delta \psi^*_{n\bm k} \hat{F}_0\psi_{n\bm k}+\psi^*_{n\bm k} \hat{F}_0\delta \psi_{n\bm k})\;.
\end{equation}
Here $\psi_{n\bm k}=e^{i\bm k\cdot \bm r} |u_n(\bm k)\rangle$ is the Bloch function of the unperturbed Hamitonian $\hat{H}$ with $\varepsilon_{n\bm k}$ being the corresponding eigenenergy, $f_{n\bm k}$ is the Fermi function, and $\hat{F}_0$ is the unperturbed part of the free energy density.  The spin toroidization can be obtained from the appropriate Fourier component of $\delta F(\bm r)$:
\begin{equation}
\mathcal{T}_z=-{2\over V h} \int dx\;\delta F(\bm r)(\cos(qx)+\cos(qy)) \;.
\end{equation}

We first calculate the contribution from the $y$-component of the Zeeman field.
The perturbation to the wave function is
\begin{equation}
\delta \psi_{n\bm k}=-{hg\mu_B\over 4i\hbar q}\Bigl[\sum_{n^\prime}{e^{i(\bm k+\bm q)\cdot \bm r}|n^\prime \bm k+\bm q\rangle \langle n^\prime \bm k+\bm q|\hat{s}_y|n\bm k\rangle\over \varepsilon_{n\bm k}-\varepsilon_{n^\prime \bm k+\bm q}}-(\bm q\rightarrow -\bm q)\Bigr] \;,
\end{equation}
where $\bm q=q \hat{x}$, and $|n\bm k\rangle$ is short for $|u_n(\bm k)\rangle$.  Inserting this expression into Eq.~\eqref{suppl_eq_free}.  We can see that the first two terms cancel each other. The last two terms read
\begin{equation} \label{eq_sup_tz}
\begin{split}
\mathcal{T}_z&={g\mu_B\over 4i\hbar q}\sum_{nn^\prime\bm k} (\varepsilon_{n\bm k}-\mu) f_{n\bm k} \left({\langle n\bm k|n^\prime \bm k+\bm q\rangle\langle n^\prime \bm k+\bm q|\hat{s}_y|n\bm k\rangle \over \varepsilon_{n\bm k}-\varepsilon_{n^\prime \bm k+\bm q}}-(\bm q\rightarrow -\bm q)\right)+\text{c.c.}\\
&={g\mu_B\over 4i\hbar q}\sum_{nn^\prime \bm k} [(\varepsilon_{n\bm k}-\mu)f_{n\bm k}-(\varepsilon_{n^\prime \bm k+\bm q}-\mu)f_{n^\prime \bm k+\bm q}]{\langle n\bm k|n^\prime \bm k+\bm q\rangle\langle n^\prime \bm k+\bm q|s_y|n\bm k\rangle-\text{c.c.} \over \varepsilon_{n\bm k}-\varepsilon_{n^\prime \bm k+\bm q}}
\end{split}
\end{equation}
Now we take the limit $q\rightarrow 0$ in the above expression.
Terms in Eq.~\eqref{eq_sup_tz} with $n\neq n^\prime$ reads
\begin{align}
\mathcal{T}_{z1}=-{g\mu_B\over 4\hbar }\sum_{n\neq n^\prime,\bm k}[(\varepsilon_{n\bm k}-\mu)f_{n\bm k}-(\varepsilon_{n^\prime \bm k}-\mu)f_{n^\prime \bm k}]{(A_x)_{nn^\prime} (\hat{s}_y)_{n^\prime n}+c.c.\over \varepsilon_{n\bm k}-\varepsilon_{n^\prime \bm k}}\,,
\end{align}
where $\bm A_{nn^\prime}=\langle n\bm k|i\bm \partial_{\bm k}|n^\prime\bm k\rangle$ is the interband Berry connection and $\bm s_{n^\prime n}=\langle n^\prime\bm k|\hat{\bm s}|n\bm k\rangle$ is the interband element of the spin operator.  Terms in Eq.~\eqref{eq_sup_tz} with $n= n^\prime$ reads

\begin{align}
\mathcal{T}_{z2}
&={g\mu_B\over 4i\hbar }\sum_{n\bm k}f_{n\bm k} (\langle \partial_x n\bm k|s_y|n\bm k\rangle+\langle n\bm k|\partial_x|n\bm k\rangle \langle n\bm k|\hat{s}_y|n\bm k\rangle-c.c.)\notag\\
&-{g\mu_B\over 4i\hbar} \sum_{n\bm k}(\varepsilon_{n\bm k}-\mu) f^\prime_{n\bm k}(\langle \partial_x n\bm k|\hat{s}_y|n\bm k\rangle-\langle n\bm k|\partial_x|n\bm k\rangle \langle n\bm k|\hat{s}_y|n\bm k\rangle-c.c.)\notag\\
&={g\mu_B\over 4\hbar}\sum_{n\neq n^\prime,\bm k} f_{n\bm k} ((A_x)_{nn^\prime}(\hat{s}_y)_{n^\prime n}+c.c.)\;.
\end{align}
Note that to obtain the last equality in the above equation, we have used the fact that at $T=0$, $f_{n\bm k}^\prime=\delta(\varepsilon_{n\bm k}-\mu)$.  The total contribution from the $y$-component of the Zeeman field thus is
\begin{align}
\mathcal{T}_z=\mathcal{T}_{z1}+\mathcal{T}_{z2}=-{g\mu_B\over 4\hbar} \sum_{n\neq n^\prime, \bm k}(\varepsilon_{n^\prime \bm k}-\mu){f_{n\bm k}-f_{n^\prime k}\over \varepsilon_{n\bm k}-\varepsilon_{n^\prime \bm k}}((A_x)_{nn^\prime}(\hat{s}_y)_{n^\prime n}+ \text{c.c.})\;.
\end{align}

We can also calculate the contribution from the $x$-component of the Zeeman field. The final result reads
\begin{equation} \label{TT}
\begin{split}
\mathcal{T}_z&=-{g\mu_B\over 4\hbar} \sum_{n\neq n^\prime, \bm k}(\varepsilon_{n^\prime \bm k}-\mu){f_{n\bm k}-f_{n^\prime k}\over \varepsilon_{n\bm k}-\varepsilon_{n^\prime \bm k}}((A_x)_{nn^\prime} (\hat{s}_y)_{n^\prime n}-(x\leftrightarrow y)+ \text{c.c.}) \\
&=-\frac{g\mu_B}{2}{\rm Im}\sum_{n\neq n^\prime,\bm k} (\varepsilon_{n^\prime\bm k}-\mu)\frac{f_{n\bm k}-f_{n^\prime \bm k}}{(\varepsilon_{n\bm k}-\varepsilon_{n^\prime \bm k})^2}(\bm v_{nn^\prime}\times \bm s_{n^\prime n})_z\;,
\end{split}
\end{equation}
where $v_{nn^\prime}=\langle n\bm k|\hat{\bm v}|n^\prime \bm k\rangle$ is the interband element of the velocity operator.  This is the multi-band formula for the spin toroidization.  It reduces to Eq.~\eqref{eq_dft} in the main text for a single band.

\section{Wannier representation}

In this section we express the spin toroidization in terms of the Wannier functions.  Denote by $|w_{0}(\bm R,\bm B)\rangle$ the Wannier function located at the lattice site $\bm R$ from band $0$, derived from the local Hamiltonian $\hat{H}_c$.  The periodic part of the Bloch function $|\tilde{u}_0\rangle$ is given by
\begin{equation} \label{app_eq_wan1}
|\tilde{u}_0\rangle={1\over \sqrt{N}}\sum_{\bm R} e^{-i\bm k\cdot(\bm r-\bm R)} |w_{0}(\bm R,\bm B)\rangle \;,
\end{equation}
where $N$ is the number of unit cells.

We begin with the spin toroidization formula in Eq.~\eqref{TT}, which can be recast as
\begin{equation}\label{app_eq_wan2}
\mathcal{\bm T}=-{g\mu_B\over 4\hbar}\sum_{n\neq 0} \int {d\bm k\over (2\pi)^3}(\bm A_{0n}\times \bm s_{n0}+\text{c.c.})-{g\mu_B\over 2\hbar}\sum_{n\neq 0}
\int {d\bm k\over (2\pi)^3}{\varepsilon_n-\mu\over \varepsilon_0-\varepsilon_n}(\bm A_{0n}\times \bm s_{n0}+ \text{c.c.})\;.
\end{equation}
The first term in Eq.~\eqref{app_eq_wan2} can be expressed in terms of $|\tilde u_0\rangle$,
\begin{align}
\sum_{n\neq 0}\bm A_{0n}\times \bm s_{n0}+c.c.=-i \langle \bm \partial_{\bm k} \tilde{u}_0|\times \hat{\bm s}|\tilde{u}_0\rangle |_{\bm B\rightarrow 0}+i \langle \bm \partial_{\bm k}\tilde{u}_0|\tilde{u}_0\rangle |_{\bm B\rightarrow 0}\times \langle \tilde{u}_0|\hat{\bm s}|\tilde{u}_0\rangle |_{\bm B\rightarrow 0}+ \text{c.c.}\;,
\end{align}
where we have used the identity: $|\tilde{u}_0\rangle\langle \tilde{u}_0|+\sum_{n\neq 0}|\tilde{u}_n\rangle\langle \tilde{u}_n|=I$.  Inserting Eq.~\eqref{app_eq_wan1} into the above expression yields

\begin{equation}\label{eq_res1}
\begin{split}
\mathcal{\bm T}_1=&-{g\mu_B\over 4N\hbar}\sum_{\bm R,\bm R_1} \int {d\bm k\over (2\pi)^3} e^{i\bm k\cdot (\bm R_1-\bm R)}\langle w_{0}(\bm R,\bm B)|(\bm r-\bm R)\times \hat{\bm s}|w_{0}(\bm R_1,\bm B)\rangle |_{\bm B\rightarrow 0}+ \text{c.c.}\\
&+{g\mu_B\over 4N^2\hbar}\sum_{\bm R,\bm R_1,\bm R_2,\bm R_3} \int {d\bm k\over (2\pi)^3} e^{i\bm k\cdot (\bm R_1-\bm R+\bm R_3-\bm R_2)} \langle w_{0}(\bm R,\bm B)|\bm r-\bm R|w_{0}(\bm R_1,\bm B)\rangle |_{\bm B\rightarrow 0}\\
=&-{g\mu_B\over 2 \hbar V_{\rm cell}}\langle w_{0}(\bm B)|\bm r\times \hat{\bm s}|w_{0}(\bm B)\rangle |_{\bm B\rightarrow 0}\\
&+{g\mu_B\over 2 \hbar V_{\rm cell}}\sum_{\bm R_1} \langle w_{0}(\bm B)|\bm r|w_{0}(\bm R_1,\bm B)\rangle |_{\bm B\rightarrow 0}\times \langle w_{0}(\bm R_1,\bm B)|\hat{\bm s}|w_{0}(\bm B)\rangle |_{\bm B\rightarrow 0}\;.
\end{split}
\end{equation}
Here $|w_0(\bm B)\rangle=|w_0(\bm R,\bm B)\rangle$ with $\bm R=0$.

Next we turn to the second term of Eq.~\eqref{app_eq_wan2}. Using $-g\mu_B \hat{\bm s}/\hbar=\bm \partial_{\bm B} \hat{H}_c$, we have,
\begin{align}
-{g\mu_B}\bm s_{n0}/\hbar &=\langle \tilde{u}_n|\bm \partial_{\bm B} \hat{H}_c|\tilde{u}_0\rangle |_{\bm B\rightarrow 0}= (\tilde{\varepsilon}_0-\tilde{\varepsilon}_n)|_{\bm B\rightarrow 0}\langle \tilde{u}_n|\bm \partial_{\bm B}\tilde{u}_0\rangle |_{\bm B\rightarrow 0}\;,
\end{align}
and
\begin{align}
&\quad-\sum_{n\neq 0}{g\mu_B\over \hbar}{\varepsilon_n-\mu\over \varepsilon_0-\varepsilon_n}(\bm A_{0n}\times \bm s_{n0}+\text{c.c.})=-i\langle \bm \partial_{\bm k}\tilde{u}_0|\times (\hat{H}_c-\mu)\bm \partial_{\bm B}|\tilde{u}_0\rangle |_{\bm B\rightarrow 0}+\text{c.c.}\;.
\end{align}
The second term in Eq.~\eqref{app_eq_wan2} then becomes
\begin{equation}\label{eq_res2}
\begin{split}
\mathcal{\bm T}_2&={1\over 2}\int {d\bm k\over (2\pi)^3}(-i\langle \bm \partial_{\bm k}\tilde{u}_0|\times (\hat{H}_c-\mu)\bm \partial_{\bm B}|\tilde{u}_0\rangle+c.c.) |_{\bm B\rightarrow 0}\notag\\
&={1\over V_{\rm cell}} {\rm Re}\langle w_0(\bm B)|\bm r\times (\hat{H}_c-\mu)\bm \partial_{\bm B} |w_0(\bm B)\rangle |_{\bm B\rightarrow 0}\;.
\end{split}
\end{equation}

Combining Eq.~\eqref{eq_res1} and Eq.~\eqref{eq_res2}, we obtain the final expression in Eq.~\eqref{eq_wan} fo the spin toroidization in the Wannier representation.

\section{Molecular insulator limit}

Under this limit the first two terms in Eq.~\eqref{eq_wan} becomes:
\begin{equation}
\mathcal{\bm T}_1 =-{g\mu_B\over 2\hbar V_{\rm cell}} \langle w_0(\bm B)|(\bm r-\bar{\bm r})\times (\hat{\bm s}-\bar{\bm s})|w_0(\bm B)\rangle |_{\bm B\rightarrow 0}\;,
\end{equation}
where $\bar{\bm r}=\langle w_0(\bm B)|\bm r|w_0(\bm B)\rangle$ is the expectation value of the position, and $\bar{\bm s}=\langle w_0(\bm B)|\hat{\bm s}|w_0(\bm B)\rangle$ is the expectation value of the spin. Since the combined time reversal and space inversion symmetry is respected, we must have $\bar{\bm s}=0$. Therefore,
\begin{equation}\label{eq_firstT}
\mathcal{T}_1 =-{g\mu_B\over 2\hbar V_{\rm cell}} \langle w_0(\bm B)|\bm r\times \hat{\bm s}|w_0(\bm B)\rangle |_{\bm B\rightarrow 0}\;.
\end{equation}

Now we consider the remaining term  in Eq.~\eqref{eq_wan}. In the molecular insulator limit, its form does not change. Note that $|w_0(\bm B)\rangle$ and $|w_n(\bm B)\rangle$ becomes the molecular eigenfunctions and $\varepsilon_0$ and $\varepsilon_n$ become the molecular eigenenergy. We further manipulate this term as follows:
\begin{equation}
\begin{split}
\mathcal{T}_2 &={1\over V_{\rm cell}}{\rm Re}\langle w_0(\bm B)|\bm r (\hat{H}_c-\mu)\times \bm \partial_{\bm B}|w_0(\bm B)\rangle |_{\bm B\rightarrow 0}\\
&={g\mu_B\over \hbar V_{\rm cell}}\sum_{n\neq 0}[\langle w_0(\bm B)|\bm r|w_n(\bm B)\rangle\times \langle w_n(\bm B)|\hat{\bm s}|w_0(\bm B)\rangle]\big |_{\bm B\rightarrow 0}-{1\over 2V_{\rm cell}} (\varepsilon_0-\mu)(\partial_{\bm B}\times \bar{\bm r})\big |_{\bm B\rightarrow 0}\\
&={g\mu_B\over \hbar V_{\rm cell}}\langle w_0(\bm B)|\bm r\times \hat{\bm s}|w_0(\bm B)\rangle\big |_{\bm B\rightarrow 0}-{1\over 2V_{\rm cell}} (\varepsilon_0-\mu)(\partial_{\bm B}\times \bar{\bm r})\big |_{\bm B\rightarrow 0}\;.
\end{split}
\end{equation}
where $\bar{\bm r}$ has been defined before, and stands for the position of electron under external magnetic field. Here $\varepsilon_0-\mu$ is the free energy for state $0$.

Therefore, the total toroidization in the molecular insulator limit reads:
\begin{equation}
\mathcal{T}=\mathcal{T}_1+\mathcal{T}_2
={g\mu_B\over 2\hbar V_{\rm cell}}\langle w_0(\bm B)|\bm r\times \hat{\bm s}|w_0(\bm B)\rangle\big |_{\bm B\rightarrow 0}-{1\over 2V_{\rm cell}} (\varepsilon_0-\mu)(\partial_{\bm B}\times \bar{\bm r})\big |_{\bm B\rightarrow 0} \;.
\end{equation}
\end{widetext}

\bibliographystyle{apsrev4-1}

\begin{thebibliography}{100}
\bibitem{foot1} The orbital contribution is more complicated. It can be defined separately from the spin part, and will be discussed elsewhere.
\end{thebibliography}

\end{document}